\documentclass[12pt]{article}

\usepackage{graphicx}%
\usepackage{amsmath,amssymb,amsfonts}%
\usepackage{amsthm}%
\usepackage{mathrsfs}%
\usepackage{xcolor}%
\usepackage{url}
\usepackage{bm}
\usepackage{a4}
\evensidemargin \oddsidemargin
\theoremstyle{thmstyleone}%
\newtheorem{theorem}{Theorem}
\newtheorem{proposition}[theorem]{Proposition}%
\newtheorem{corollary}[theorem]{Corollary}%

\theoremstyle{thmstyletwo}%

\theoremstyle{thmstylethree}%

\newcommand{\NN}{\mathbb{N}}

\def\1{{\bf 1}}
\def\0{{\bf 0}}

\def \D {{\cal D}}

\def \I {{\cal I}}

\def \R {{\cal R}}

\def \U {{\cal U}}

\newcommand{\Be}{{\bm \epsilon}}
\newcommand{\Bep}{{\bm \epsilon}^{\scriptscriptstyle \perp}}

\newcommand{\bi}{\mathbf{i}}
\newcommand{\bj}{\mathbf{j}}
\newcommand{\bk}{\mathbf{k}}

\def\nn{\nonumber \\}

\newcommand{\bx}{{\bm x}}

\newcommand{\bX}{{\bm X}}
\newcommand{\bXp}{{\bm X}^{\scriptscriptstyle \perp}}

\newcommand{\bv}{{\bm v}}

\newcommand{\Bap}{{\bm \alpha}^{\scriptscriptstyle \perp}}

\newcommand{\bE}{{\bm E}}
\newcommand{\bEp}{{\bm E}^{\scriptscriptstyle \perp}}
\newcommand{\bB}{{\bm B}}

\newcommand{\hD}{\hat \Delta}
\newcommand{\hs}{\hat  s}
\newcommand{\hH}{\hat  H}
\newcommand{\hJ}{\hat  J}

\newcommand{\hbx}{\hat \bx}
\newcommand{\hxe}{\hat x_e}
\newcommand{\hye}{\hat y_e}
\newcommand{\hze}{\hat z_e}
\newcommand{\hp}{\hat  p}
\newcommand{\hbp}{\hat {\bm p}}
\newcommand{\hbu}{\hat {\bm u}}
\newcommand{\hDO}{\Delta\!^{{\scriptscriptstyle (0)}}}
\newcommand{\hDU}{\Delta\!^{{\scriptscriptstyle (1)}}}

\newcommand{\hDOu}{\Delta_u}
\newcommand{\hDd}{\Delta_d}

\newcommand{\vM}{v_{{\scriptscriptstyle M}}}
\newcommand{\tvM}{\tilde v_{{\scriptscriptstyle M}}}
\newcommand{\xiH}{\xi_{{\scriptscriptstyle H}}}

\newcommand{\sU}{s^{{\scriptscriptstyle (1)}}}
\newcommand{\sD}{s^{{\scriptscriptstyle (2)}}}
\newcommand{\hsU}{\hs ^{{\scriptscriptstyle (1)}}}
\newcommand{\hsD}{\hs ^{{\scriptscriptstyle (2)}}}

\newcommand{\be}{\begin{equation}}
\newcommand{\ee}{\end{equation}}
\newcommand{\bea}{\begin{eqnarray}}
\newcommand{\eea}{\end{eqnarray}}
\newcommand{\ba}{\begin{array}}
\newcommand{\ea}{\end{array}}

\begin{document}

\title{Hydrodynamic regime and cold plasmas hit by short laser pulses}

\author{Gaetano Fiore$^{1,3}$, \ Monica De Angelis$^{1}$, Renato Fedele$^{2,3}$, \\
 Gabriele Guerriero$^{1}$, Du\v{s}an Jovanovi\'c$^{4,5}$, 
   \\    
$^{1}$ Dip. di Matematica e Applicazioni, Universit\`a di Napoli ``Federico II'',\\
$^{2}$  Dip. di Fisica, Universit\`a di Napoli ``Federico II'', \\
Complesso Universitario  M. S. Angelo, Via Cintia, 80126 Napoli, Italy\\         
$^{3}$         INFN, Sez. di Napoli, Complesso  MSA,  Via Cintia, 80126 Napoli, Italy\\ 
$^{4}$  Inst. of Physics, University of Belgrade, 
11080 Belgrade, Serbia\\
$^{5}$  Texas A \& M University at Qatar, 23874 Doha, Qatar 
}

\date{}
\maketitle

\begin{abstract} 
We briefly report and elaborate on some conditions allowing a hydrodynamic description of the impact of  a  very short and arbitrarily intense laser pulse onto a cold plasma, as well as  the localization of the first wave-breaking due to the plasma inhomogeneity. 
We use a  recently developed fully relativistic plane model whereby we reduce the system of the Lorentz-Maxwell and continuity PDEs  into a 1-parameter
 family of decoupled systems of  non-autonomous Hamilton equations in dimension 1,
 with the light-like coordinate $\xi=ct\!-\!z$ replacing time $t$ as an independent variable.
Apriori estimates on the Jacobian $\hat J$ of the change from Lagrangian to Eulerian coordinates in terms of the input data (initial density and pulse profile) are obtained applying Liapunov direct method to an associated family of pairs of ODEs; wave-breaking is pinpointed by the inequality $\hat J\le 0$. 
These results may help in drastically simplifying the study of extreme acceleration mechanisms of electrons, which have very important applications.
\end{abstract}



\section{Introduction and plane model}\label{sec1}

Ultraintense 
laser-plasma interactions lead to  exciting phenomena \cite{Kruer19,SprEsaTin90PRA,SprEsaTin90PRL,EsaSchLee09,Mac13}, 
notably plasma compression for inertial fusion \cite{KuzRyz17},  laser wakefield acceleration  (LWFA)  \cite{Tajima-Dawson1979,Sprangle1988,TajNakMou17} and other extremely compact acceleration mechanisms
 of charged particles, which hopefully will allow the production of new, table-top accelerators. 
Huge investments are presently devoted to the development of the latter\footnote{We just mention the EU-funded project {\it Eupraxia} \cite{Eupraxia19AIP,Eupraxia19JPCS,Eupraxia20EPJ}.}, because their small size
would drastically  facilitate the extremely important applications of accelerators in particle physics, medicine, material science, industry, inertial fusion,  environmental remediation, 
etc. 
In general, these phenomena are ruled by the equations of a relativistic kinetic theory coupled to Maxwell equations, which today can be  solved numerically via increasingly powerful particle-in-cell (PIC) codes. However, since the simulations involve huge costs for each choice of the input  data, exploring the data space blindly to single out interesting regions remains prohibitive. All analytical insights that can simplify
the work, at least in special cases or in a limited space-time region, are welcome.
Sometimes,  good predictions can be obtained also by a hydrodynamic description  (HD) of the plasma, i.e. treating it as a multicomponent (electron and ions) fluid, and by numerically solving the (simpler) associated hydrodynamic equations via  multifluid (such as QFluid \cite{TomEtAl17}) or hybrid kinetic/fluid  codes; but in general it is not known a priori
in which conditions, or spacetime regions, this is possible. 

Here we summarize  and slightly elaborate on a set of conditions  \cite{FioDeAFedGueJov22}
enabling a rather simple HD   of the impact of a very short (and possibly very intense)  laser pulse onto a cold diluted plasma at rest and the localization  {\it after} the impact
of the first wave-breakings (WBs) of the plasma wave (PW) \cite{AkhPol56,GorKir1987} due to  inhomogeneities of the initial density  \cite{Daw59}.
Our analysis is based on a fully relativistic plane Lagrangian model   \cite{Fio14JPA,Fio18JPA,FioCat18} and very little computational power.
We recall that small WBs are not necessarily undesired  where the initial density decreases: they may be used  \cite{BulEtAl98} to inject and trap a small bunch of plasma electrons as {\it test  electrons} in the PW trailing the pulse  ({\it self-injection}), so that these undergo LWFA  in the forward direction.
The impact of very short laser pulses on suitable initial plasma profiles may allow  also the {\it slingshot effect}  \cite{FioFedDeA14,FioDeN16,FioDeN16b},
i.e. the backward acceleration and expulsion of (less) energetic electrons from the vacuum-plasma interface, during or just after the impact.

\begin{figure}
\includegraphics[height=3.375cm]{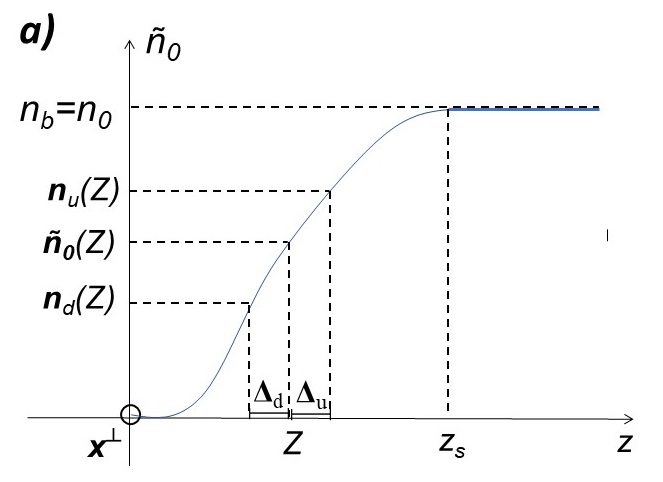}\includegraphics[height=3.375cm]{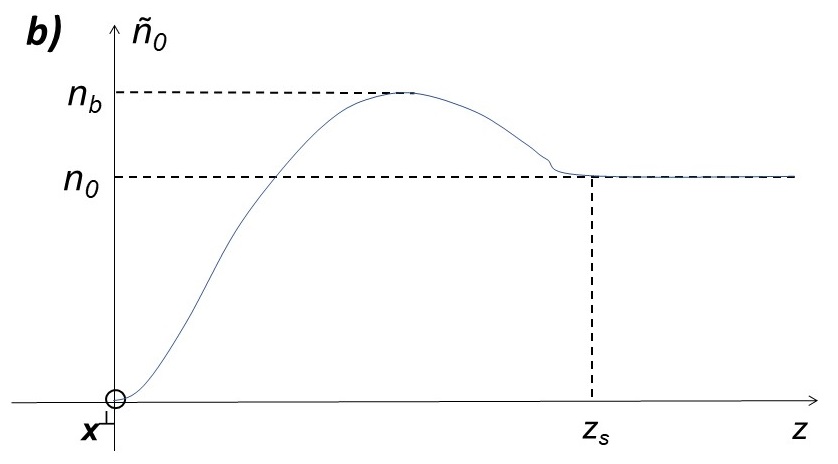}
\includegraphics[height=3.45cm]{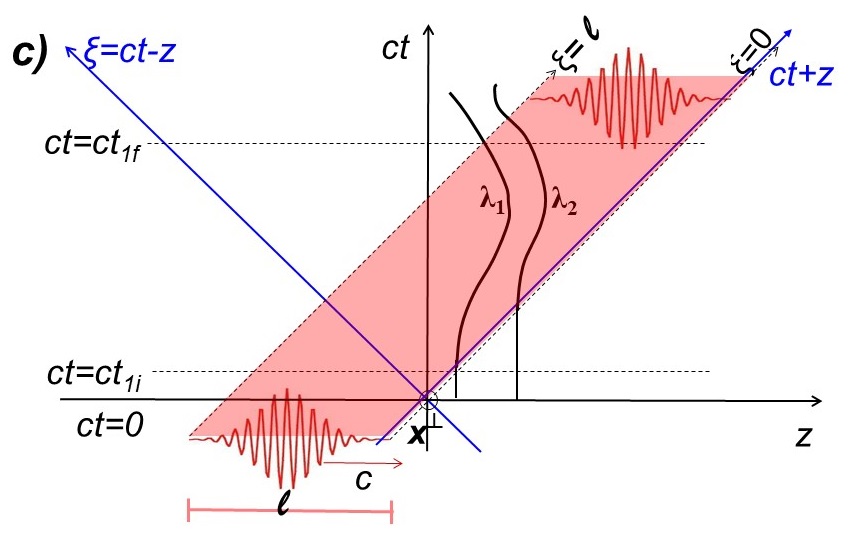}
\caption{a), b): Examples of initial plasma densities of the type
(\ref{n_0bounds}).  In a) we also illustrate the meaning of the functions $n_u(z),n_d(z)$ defined in (\ref{n-bounds}).
c): Projections onto the $z,ct$ plane of sample particle worldlines (WLs) $\lambda_1,\lambda_2$  in Minkowski space \cite{FioCat18}; they intersect the support (pink) of a plane EM wave of total length $l$ moving in the positive $z$ direction. 
Since each WL intersects once every hyperplane $\xi\!=$ const (beside every hyperplane $t\!=$ const), we can use $\xi$ rather than $t$ as a parameter along it. While the $t$-instants of intersection with the front and the end of the EM wave (e.g. $t_{1i},t_{1f}$ for $\lambda_1$) depend on the particular WL, the corresponding $\xi$-intstants are the same for all WLs: $\xi_i=0$, $\xi_f=l$.}
\label{fig1}
\end{figure}

The plane model is as follows. 
One assumes that the plasma is initially neutral, unmagnetized and at rest with zero densities in the region  \ $z\!<\! 0$.
More precisely, the $t\!=\!0$ initial conditions for  the electron fluid  Eulerian density $n_e$ and velocity $\bv_e$ are  
\be 
\bv_e(0 ,\!\bx)\!=\!\0, \qquad n_e(0,\!\bx)\!=\!\widetilde{n_0}(z), 
 \label{asyc}
\ee
where the initial electron (as well as proton) density $\widetilde{n_0}(z)$ fulfills
\be 
 \widetilde{n_0}(z)\!=\!0 \:\:\mbox{if }\: z\!\le\! 0, \qquad
0\!<\!\widetilde{n_0}(z)\!\le\! n_b  \quad \mbox{if }\: z\!>\!0
 \label{n_0bounds}
\ee
for some $n_b\!>\!0$  (two examples of such densities are reported in fig. 	\ref{fig1}). 
One assumes that before the impact the laser pulse  is a free plane transverse wave  
 travelling  in the $z$-direction, i.e.
the electric and magnetic fields $\bE,\bB$ are of the form
\be
\bE (t, \bx)=\bEp (t, \bx)=\Be^{{\scriptscriptstyle\perp}}\!(ct\!-\!z),\qquad  \bB=\bB^{{\scriptscriptstyle\perp}}=
\bk\!\times\!\bEp \qquad \mbox{if } t\le 0             
      \label{pump}
\ee
(given a vector ${\bm w}$, we denote by ${\bm w}^\perp$ its component $\perp\bk\equiv\nabla z$), where the support of
 $\Be^{{\scriptscriptstyle\perp}}\!(\xi)$ is  
a suitable interval $[0,l]$
($\xi\!=\!0$ as the left extreme means that the pulse reaches the plasma at $t\!=\!0$; $l$ is constrained below). The input data of a specific problem are the functions $\widetilde{n_0}(z),\Be^{{\scriptscriptstyle\perp}}(\xi)$; 
it is useful to introduce also the related functions
\begin{align}
 \displaystyle \Bap(\xi)
\equiv -\!\int^{\xi}_{ -\infty }\!\!\!d\zeta\:\Bep(\zeta), \qquad
\qquad\qquad\quad\:\:
v(\xi)
\equiv\left[\frac {e\Bap(\xi)}{mc^2}\right]^2,
  &    \label{pump2}\\
 \displaystyle \widetilde{N}(Z)
\equiv\int^{Z}_0\!\!\! d \zeta\,\widetilde{n_0}(\zeta), \qquad
\U( \Delta;Z)
\equiv  K\!\!\int^{\Delta}_0\!\!\!\!\!d\zeta\,(\Delta\!-\!\zeta)
\,\widetilde{n_0}(Z\!+\! \zeta)\:; &
\end{align}
$-e,m$ are the electron charge and mass, $c$ is the speed of light,  $K\equiv\frac{4\pi e^2}{mc^2}$.  By definition, $v$ is dimensionless and nonnegative, $\widetilde{N}(z)$ strictly grows with $z$.
When reached by the pulse, electrons start oscillating transversely (i.e. in the $x,y$ directions) 
and drifting in the positive $z$-direction, respectively pushed by  the electric and magnetic parts of the Lorentz force due to  the pulse; thereafter, electrons  start oscillating also longitudinally (i.e. in $z$-direction), pushed by the restoring electric force due to charge 
separation. 
We shall  assume that the length $l$ of the pulse makes the latter {\it essentially short} (ES) w.r.t. the  density $\widetilde{n_0}$, in the sense of definition (\ref{Lncond'}),
implying that the pulse overcomes each electron before the $z$-displacement $\hD$ of the latter  reaches a negative minimum for the first time.
Most applications use slowly modulated monochromatic
(SMM)  waves
\be
\Bep\!(\xi)\!=\!\underbrace{\epsilon(\xi)}_{\mbox{modulation}}
\underbrace{[\bi \cos\psi\,\sin (k\xi\!+\!\varphi_1)\!+\!\bj \sin\psi\sin (k\xi\!+\!\varphi_2)]}_{\mbox{carrier wave $\Be_o^{{\scriptscriptstyle \perp}}\!(\xi)$}},
 \label{modulate}
\ee
where $\bi\!=\!\nabla x$,  $\bj\!=\!\nabla y$, and the length $\lambda\!=\!2\pi/k$ of the carrier wave is much smaller than the length $l$ of
the support $[0,l]$ of $\epsilon(\xi)$.  
Then  \ $\Bap(\xi)= -\Bep_o(\xi\!+\!\lambda/4)\: \epsilon(\xi)/k$ \ up to terms $O\big((\lambda/l)^2\big)$  (see appendix 5.4 in \cite{Fio18JPA} for  details), whence $\Bap(\xi),v(\xi)\simeq 0$ for $\xi\ge l$.
As we recall below, if $v(\xi)\ll 1$ for all $\xi$ then electrons keep nonrelativistic (NR); by Proposition 1 of \cite{FioDeAFedGueJov22}, the pulse is  ES if 
the modulation is symmetric about its center $\xi=l/2$ (i.e. $\epsilon(\xi)=\epsilon(l\!-\!\xi)$) and its duration $l/c$  does not exceed the NR plasma oscillation period $t_{{\scriptscriptstyle H}}^{{\scriptscriptstyle nr}}\!\equiv\!\sqrt{\pi m/n_b e^2}$ associated to the maximum $n_b$ of $ \widetilde{n_0}(z)$, i.e. if
\be
G_b\equiv\sqrt{\frac{n_b e^2}{\pi mc^2}}l \: \le \: 1
   \label{Lncond}
\ee
(whence  $\frac{4\pi e^2}{mc^2}n_b \lambda^2\ll1$, and the plasma is {\it underdense}).
A general sufficient condition \cite{FioDeAFedGueJov22} for a pulse to be ES will be recalled in  formula (\ref{ShortPulse1'}) below; it may be $G_b>1$.

One describes the plasma as a fully relativistic collisionless fluid of electrons and a static fluid of ions (as usual, in the short time lapse of interest here the motion of the much heavier ions is negligible), 
with  $\bE,\bB$ and the plasma dynamic variables fulfilling the Lorentz-Maxwell and continuity equations. Since at the impact time $t\!=\!0$ the plasma is made of two static fluids, by continuity such a hydrodynamic description (HD) is justified and  one can neglect  the depletion of the pulse  at least for small $t\!>\!0$;  the specific time lapse is 
determined   {\it a posteriori},  by self-consistency (see e.g. \cite{FioDeNAkhFedJov23}).  
This allows us to  reduce (see \cite{Fio14JPA,Fio18JPA}, or \cite{Fio14,Fio16b,FioCat19,Fio21JPCS} for
 shorter presentations) the system of Lorentz-Maxwell and continuity 
partial differential equations (PDEs)  into ordinary ones, 
more precisely into a continuous family of   {\it decoupled Hamilton equations for  systems  with one degree of freedom}. Each system rules the Lagrangian  (in the sense of non-Eulerian) description of the motion of the electrons having a same
initial longitudinal coordinate $Z>0$ (the {\it $Z$ electrons}, for brevity), and reads
\bea
&\hD '(\xi,Z)
=\displaystyle\frac {1\!+\!v(\xi)}{2\hs ^2(\xi,Z)}\!-\!\frac 12, \qquad 
&\hs '(\xi,Z) 
=K\!\left\{\!\widetilde{N}\left[Z\!+\!\hD (\xi,Z)\right] \!-\! \widetilde{N}(Z)\!\right\};
\label{heq1}
\eea
it is  equipped with the initial conditions
\bea
 &\hD (0,Z)=0, \qquad\quad &\hs (0,Z)=1.               \label{incond}
\eea
Here  the unknowns $\hD (\xi,Z),\hs (\xi,Z)$  are respectivey  the
present longitudinal displacement and $s$-factor\footnote{Namely, $\hs $ is the light-like component of the 4-velocity  of the $Z$ electrons, or equivalently  is related to their 4-momentum $\hp$
by $\hp^0\!-\!c\hp^z\equiv mc^2 \hs $; it is positive-definite. In the NR regime $|\hs\!-\!1|\ll1$; in the present fully relativistic regime it needs only satisfy the inequality $\hs>0$.} of the $Z$ electrons espressed as functions of
 $\xi,Z$, while  $\hze(\xi,Z)
\equiv Z\!+\!\hD (\xi,Z)$ is  the present longitudinal coordinate of the $Z$ electrons; 
we express all dynamic variables $\tilde f(t,Z)$ (in the Lagrangian description)  
as functions $\hat f$ of $\xi,Z$; $\hat f'$ stands for the total derivative 
$d\hat f/d\xi\equiv \partial \hat f/\partial \xi\!+\!\hs'\partial \hat f/\partial \hs\!+\!\hD'\partial \hat f/\partial \hD$; $Z$ plays the role of the family parameter. 
The  light-like coordinate 
$\xi=ct\!-\!z$  in Minkowski spacetime can be 
adopted   instead of time $t$ as an independent variable because all particles
must travel at a speed lower than $c$, see fig. \ref{fig1}.c; at the end, to express the solution as a function of $t,Z$ one just needs to replace everywhere $\xi$ by the inverses 
$\tilde\xi(t,Z)$ of the strictly increasing (in $\xi$) functions
$\hat t(\xi,Z)\equiv (\xi\!+\!\hze(\xi,Z))/c$, with $Z\ge 0$. All the electron dynamic
variables can be expressed in terms of the basic ones $\hD ,\hs $ and the initial
coordinates $\bX\equiv(X,Y,Z)$ of the generic electron fluid element.
In particular,   the electrons' transverse momentum in $mc$ units
is given by   $\hbu^{\scriptscriptstyle \perp}=\hbp^{\scriptscriptstyle \perp}/mc=\frac {e\Bap }{mc^2}$, and $v=\hbu^{\scriptscriptstyle \perp}{}^2$.   Ultra-intense pulses are characterized by
$\max_{\xi\in[0,l]}\{v(\xi)\}\gg 1$ and induce ultra-relativistic electron motions.
  Eq. (\ref{heq1}) are Hamilton equations with $\xi,\hD , -\hs $  playing the role of the usual $t,q,p$ and  (dimensionless) Hamiltonian
\bea
\check H( \hD , \hs,\xi;Z)\equiv  \frac{\hs^2+1\!+\!v(\xi)}{2\hs}
+ \U( \hD ;Z);  
                               \label{hamiltonian}
\eea
the first term gives the electron relativistic factor $\hat\gamma$, while $\U$ plays the role of a potential energy due to the electric charges' mutual interaction. 
 For $\xi\ge l$ eqs (\ref{heq1}) are autonomous   and can be solved also by quadrature, since  the Hamiltonian $\hH(\xi,Z)\equiv \check H\big[\hD(\xi,Z), \hs(\xi,Z),\xi;Z\big]$ becomes $h(Z)\equiv\hH(l,Z)=$ const. The solutions of (\ref{heq1}-\ref{incond})
yield the motions of the $Z$ electrons'  fluid elements, which are fully represented by their worldlines (WLs) in Minkowski space.
In fig.  \ref{Worldlines_HomLin-Nonlin}  we display
the projections onto the $z,ct$ plane of a set of WLs for two specific sets of input data; as evident, the PW emerges from them as a collective effect. 
Mathematically, the PW features are derived passing to the Eulerian description of the electron fluid; the resulting flow is laminar with $xy$ plane symmetry.
The Jacobian  of the transformation 
$\bX\mapsto\hbx_e\equiv(\hxe,\hye,\hze)$ 
 from the Lagrangian  to the Eulerian coordinates
reduces to $\hJ(\xi,Z)=\partial\hze(\xi,Z)/\partial Z$, 
because  $\hbx^{\scriptscriptstyle \perp}_e\!-\!\bXp$ does not depend on $\bXp$.
The HD breaks where WLs intersect, leading to WB of the PW. 
No WB occurs as long as  $\hJ>0$ for all $Z\ge 0$.
If the initial density is uniform, then
(\ref{heq1}-\ref{incond}), and hence  also their solutions, are $Z$-independent,
and $\hJ\equiv 1$ for all $\xi,Z$. Otherwise, WB   occurs after a sufficiently long time \cite{Daw59}.

In section \ref{DsHbounds} we present upper and lower bounds  on
$\hs ,\hD$  \cite{FioDeAFedGueJov22} that provide useful approximations of these dynamic variables in the interval $0\le\xi\le l$. 
In section \ref{Jbound}  
we use these bounds to formulate sufficient conditions on the input data $\widetilde{n_0}(z),\Be^{{\scriptscriptstyle\perp}}(\xi)$  guaranteeing that $\hJ(\xi,Z)>0$ 
for all $Z>0$ and $\xi\in[0,l]$, so that there is no wave-breaking during the laser-plasma interaction (WBDLPI). These conditions are derived  \cite{FioDeAFedGueJov22} with the help of a suitable Liapunov function and now can be more easily checked where $\widetilde{n_0}$ is concave, thanks to the new results of Proposition \ref{prop1} and Corollary \ref{coroll}.
Qualitatively, $\widetilde{n_0}(z)$ and/or its local relative variations must be sufficiently small. 
For $\xi\ge l$, while   $\hD$ and $\hs$ are periodic with a suitable period $\xiH$,  $\hJ$ satisfies \cite{FioDeNAkhFedJov23} (section \ref{Jbound})
\be
\hJ(\xi,Z)=a(\xi,Z)+\xi \, b(\xi,Z), \qquad \xi\ge l,
 \label{lin-pseudoper}
\ee
where $a,b$ are periodic in $\xi$ with period $\xiH(Z)$, and $b$  has zero average over  a period.
As $b$  oscillates between positive and negative values, so does the second term,
 which dominates as $\xi\to \infty$, with $\xi$ acting as a modulating amplitude.
Localizing WBs {\it after} the laser-plasma interaction is best
investigated via 
 (\ref{lin-pseudoper}) \cite{FioDeNAkhFedJov23}. 
In section \ref{discuss} we briefly compare the dynamics of $\hs,\hD,\hJ
$ induced by the same pulse on two different $\widetilde{n_0}$s having the same  upper bound $n_b$.
Their behaviour for $z\simeq 0$ is crucial; 
WBDLPI can be excluded under rather broad conditions for typical LWFA experiments.
We also comment on the spacetime region $\R$ where the model's predictions are reliable.
 Other typical phenomena of plasma physics (turbulent flows, diffusion, heating, 
moving ions,...) can be excluded  inside $\R$, but can and will occur outside.

\section{Apriori estimates of  $\hD ,\hs
$ for small $\xi>0$}  
\label{DsHbounds}

\begin{figure}
\includegraphics[width=17.685cm]{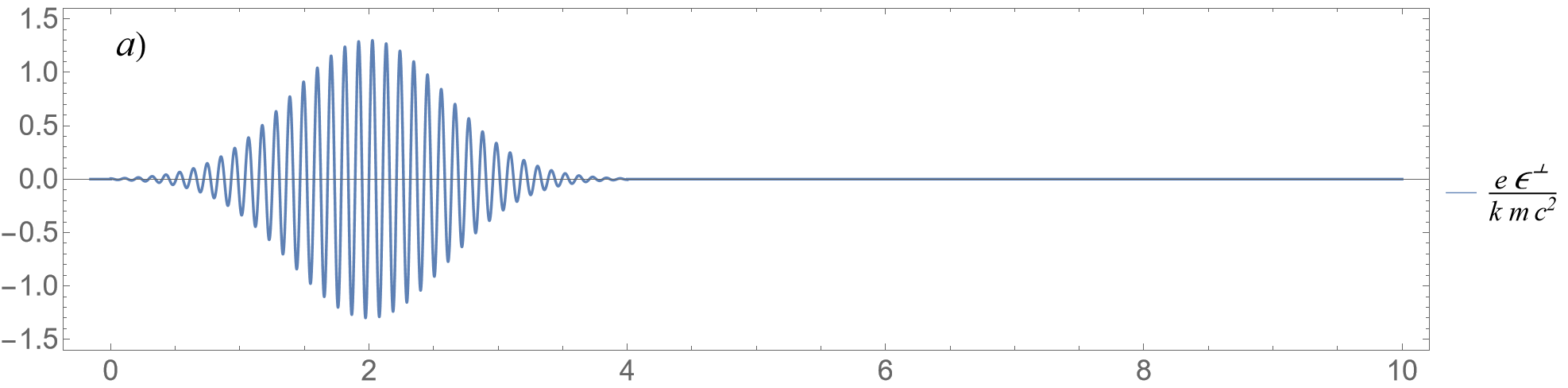}\\
\includegraphics[width=17.55cm]{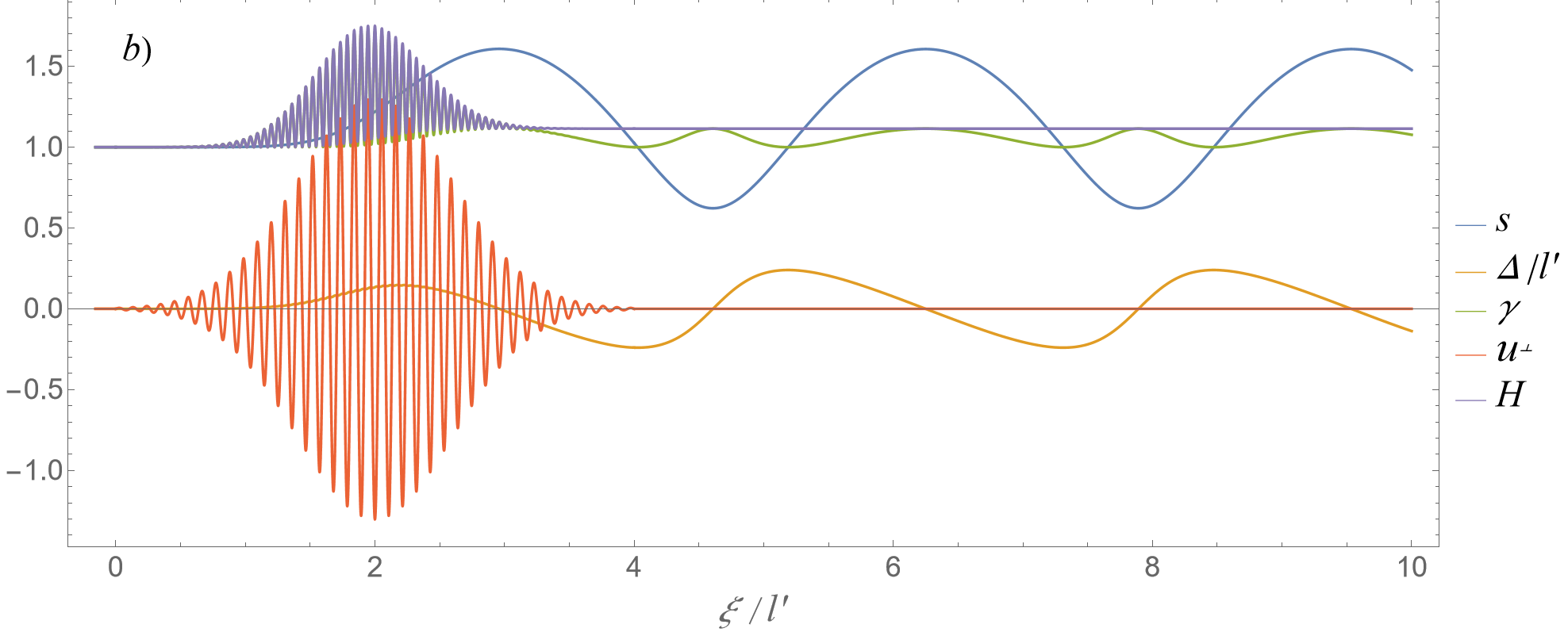} 
\caption{\ a)  Normalized amplitude of a  linearly polarized [$\psi=0$ in (\ref{modulate})]
SMM laser pulse, modulated by a Gaussian with {\it full width at half maximum} $l'$ and 
{\it peak  amplitude} $a_0\!\equiv\!\lambda eE^{\scriptscriptstyle \perp}_{\scriptscriptstyle M}/mc^2\!=\!1.3$; this makes electrons moderately relativistic, and $\Delta_u\equiv\hDO(l)\simeq 0.45 l'$. The pulse is ES w.r.t. $\widetilde{n_0}(Z)\!\equiv\!n_0\!=\!4/Kl'{}^2$.  
\ \ b) \ 
Corresponding solution of (\ref{heq1}-\ref{incond}).
As expected: $\hs$ is insensitive to the rapid oscillations of $\Bep$; 
 for $\xi\ge l$ the energy $\hH$ is conserved, and the solution is periodic. The pulse length $l$ is determined
on physical grounds; if e.g. the plasma is created locally by the impact  of the front of the
pulse   on a gas  (e.g. hydrogen or helium), then $[0,l]$ consists of all points $\xi$  where the pulse intensity is sufficient to transform the gas  into a plasma by ionization. Here  instead we conventionally fix  $l=4l'$, what makes $G_b=\sqrt{Kn_0}\,l/2\pi\simeq 1.27$. 
If $l'\!=\!7.5\mu$m, 
$\lambda=0.8\mu$m, then $n_0=2\times 10^{18}$cm$^{-3}$
and the  peak  intensity is $I\!=\!7.25\!\times\!10^{18}$W/cm$^2$;
these are typical values in  LWFA experiments with Ti:Sapphire lasers.
}
\label{graphs}
\end{figure}

The $Z$-dependent Cauchy problems (\ref{heq1}-\ref{incond}) are equivalent to the following integral ones:
\be
\hD (\xi,Z)= \int_0^\xi\!\!d\eta\,\frac {1\!+\!v(\eta)}{2\hs ^2(\eta,Z)}-
\frac{\xi}2,\qquad \hs (\xi,Z)-1= \int_0^\xi\!\!\!d\eta\!\!\int^{\hze(\eta,Z)}_Z\!\!\!\!\!\!\!\!\!\!\!\!
dZ' \: K\widetilde{n_0}(Z').
  \label{sDelta}
\ee
By (\ref{heq1}b), the zeroes of $\hD(\cdot,\!Z)$ 
are extrema of $\hs(\cdot,\!Z)$, because $\widetilde{N}(Z)$ strictly grows with $Z$; conversely, 
by (\ref{heq1}a) the zeroes of $\hs^2(\cdot,\!Z)-1-v(\cdot)$ are  extrema of $\hD(\cdot,\!Z)$.
We recall how $\hD,\hs$ start evolving  from their initial values (\ref{incond}).  
As said, for small $\xi\!>\!  0$ all electrons reached by the pulse start oscillating  transversely and drifting forward; 
in fact, $v(\xi )$  becomes positive, implying in turn that so does the right-hand side (rhs) of (\ref{heq1}a) and $\hD$;   
the $Z\!=\!0$ electrons leave behind themselves a  layer of ions 
completely evacuated of electrons (see fig. \ref{Worldlines_HomLin-Nonlin}).
If the density vanished identically
 ($\widetilde{n_0}\equiv 0$) then we would obtain
$$
\hs \equiv 1, \qquad \hD (\xi,Z) 
=\int_0^\xi\!\!d\eta\,\frac {v(\eta)}{2}=:\hDO(\xi);
$$
$\hDO(\xi)$ grows with $\xi$.
Conversely, $\widetilde{n_0}> 0$, and the growth of $\hD$ implies also that of
the rhs of (\ref{heq1}b) (because the latter grows with $\hD$) and of $\hs(\xi,\!Z)\!-\!1$.  \ $ \hD(\xi,\!Z)$ keeps growing  as long as $1\!+\!v(\xi)>\hs^2(\xi,\!Z)$, reaches a maximum at $\tilde\xi_1(Z)\equiv$ the smallest  $\xi\!>\!0$ such that the rhs (\ref{heq1}a)  vanishes.
$ \hs(\xi,\!Z)$ keeps growing  as long as $\hD(\xi,\!Z)\!\ge\!0$, reaches a maximum at the first zero $\tilde\xi_2>\tilde\xi_1$ of $ \hD(\xi,\!Z)$ and  decreases  for $\xi>\tilde\xi_2$, while  $\hD(\xi,\!Z)$ is negative. $ \hD(\xi,\!Z)$  reaches a negative minimum at 
$\check\xi_3(Z)\equiv$ the smallest  $\xi\!>\!\tilde\xi_2$ such that the rhs (\ref{heq1}a)  vanishes 
again. 
We also denote by $\tilde\xi_3(Z)$ the smallest  $\xi\!>\!\check\xi_3$ such that $ \hs(\xi,\!Z)=1$.
We invite the reader to single out $\tilde\xi_1,\tilde\xi_2,\check\xi_3,\tilde\xi_3$ for the solution displayed in fig. \ref{graphs}b.
As said, if $\Bep$ is a SMM wave,  then  for $\xi>l$ we have  $v(\xi)=v(l) \simeq 0$,
$\hDO(\xi)$ is almost constant, and $\check\xi_3\simeq\tilde\xi_3$ if in addition $l<\check\xi_3$.
We shall say that 
\be
\ba{ll}
\mbox{{\it a pulse is essentially short (ES) w.r.t. $\widetilde{n_0}$} if}\quad & 
\hs(\xi,Z)\ge 1, \\[6pt]
\mbox{{\it a pulse is strictly short (SS) w.r.t. $\widetilde{n_0}$} if}\quad & 
\hD(\xi,Z)\ge 0,
\ea    \label{Lncond'}
\ee
for all $\xi\!\in\![0,\!l]$, $Z\!\ge\! 0$; \ 
equivalently, a pulse is ES (resp. SS) if $l\le \tilde\xi_3(Z)$ 
(resp. $l\le \tilde\xi_2(Z)$) for all $Z\!\ge\! 0$. Clearly,  a SS pulse is also ES. 
As we now see, ES pulses are recommendable because they allow useful apriori bounds on $\hs,\hD,\hH, \hJ$ and thus simplify the control of the PW and its WB; moreover,
a suitable ES pulse with $l\sim \tilde\xi_2(Z)$ maximizes the energy transfer from the pulse to the $Z$ electrons \cite{SprEsaTin90PRL,FioFedDeA14}.

In fact, setting \ $\tilde\xi_3'\equiv\min\{l,\tilde\xi_3\}$, \ $\check n(\xi,Z)\!\equiv\!\widetilde{n_0}\big[\hze(\xi,\!Z)\big]$, \ by Proposition 2 in \cite{FioDeAFedGueJov22}  
\bea
&& \Delta_d(Z)\le \hDU(\xi,Z)\le\hD (\xi,Z)\le\hDO(\xi)\le\Delta_u,\label{Deltabound1} \\[8pt]
&&  1\le\hsD(\xi,Z)\le\hs (\xi,Z)\le\hsU(\xi,Z)\le s_u(Z),\label{sbound1} \\[8pt]
&& n_d(Z)\le \check n(\xi,Z)
\le n_u(Z)\le n_b,
\label{nbound1}
\eea
for all \ $Z\!\ge\! 0$, \ $\xi\!\in\![0,\tilde\xi_3']$ (i.e. for all $\xi\!\in\![0,l]$, if  the pulse is ES), where we have defined
\bea
&& \hDO\!(\xi)\equiv\int_0^\xi\!\!d\eta\,\frac {v(\eta)}{2},\qquad
\hDOu\equiv\hDO\!(l),\qquad n_u''(z)\equiv\max\limits_{z\le\zeta\le z\!+\!\hDOu}\big\{\widetilde{n_0}(\zeta)\big\},\\[6pt]
&& \hDd(z) \equiv\mbox{the negative solution of the eq. } \:
\U( \Delta;\!z)=\ba{c}\frac K2\ea \hDOu^2\,n_u''(z),\\[8pt]
&& n_u(z)\equiv \max\limits_{\zeta\in \I_z} \{\widetilde{n_0}(\zeta)\},
\quad n_d(z) \equiv \min\limits_{\zeta\in \I_z} \{\widetilde{n_0}\zeta)\},
\qquad   \I_z\!\equiv\!\big[z\!+\!\hDd,z\!+\!\hDOu\big],
\label{n-bounds} \\[4pt]
&&  M_u\equiv Kn_u,\quad  M_d\equiv Kn_d,\qquad
\left.\ba{l}s_u\\ s_d\ea\! \right\}
\equiv 1\!+\!\frac{M_u}{2}\hDOu^2\!\pm\!\sqrt{\!\left(\!1\!+\!\frac{M_u}{2}\hDOu^2\!\right)^2\!-\!1}
\eea
(as a first estimate, $\hDd=-\Delta_u$; note also that \ $1/s_d=s_u>1$), and
\bea
&& \hsU(\xi,\!z)\equiv\min\left\{s_u,1 +g(\xi,\!z)\right\},\qquad
g(\xi,\!z)\equiv\frac {M_u}{2}\!\!\int_0^\xi\!\!\!\!d\eta\,(\xi\!-\!\eta)\,v(\eta),  \\[8pt]
&& \ba{l}
\displaystyle f(\xi,\!z)\equiv\!\!\int_0^\xi\!\!\! d\eta\,(\xi\!-\!\eta)\!\left(\!\frac {1\!+\!v(\eta)}{\left[\hsU(\eta,\!z)\right]^2}- 1\!\right),\quad
\tilde\xi_2^{{\scriptscriptstyle (1)}}(z)\equiv\max_{\xi\ge 0}\big\{ f(\xi,\!z) \big\}
<\tilde\xi_2(z)  \\[14pt]
\displaystyle  M_u'=Kn_u', \qquad   \qquad 
n_u'(z)\equiv\max\limits_{z+\hDd \le\zeta\le z}\big\{\widetilde{n_0}(\zeta)\big\}\:\le \:n_u(z), 
\\[10pt]
\displaystyle  
\hsD(\xi,z)\equiv \left\{\!  \ba{ll}
1+ \frac{M_d}2  f(\xi,z) \quad & 0\le\xi\le \tilde\xi_2^{{\scriptscriptstyle (1)}},\\[8pt]
\max\left\{s_d,1\!+\! \left(\!\frac{M_d}2\!-\! \frac{M_u'}2\!\right) \! f\big(\tilde\xi_2^{{\scriptscriptstyle (1)}}\!,\!z\big) \!+\!
\frac{M_u'}2 f(\xi,\!z)\right\}\quad & \tilde\xi_2^{{\scriptscriptstyle (1)}}< \xi\le \tilde\xi_3',\ea\right.
\ea \qquad  \\[8pt]
&& \ba{l}
\displaystyle  \hDU(\xi,\!z)\equiv\max\left\{\hDd,d(\xi,\!z)\right\},\quad d(\xi,\!z)\equiv
\!\!\int_0^\xi\!\!\!\!d\eta\,\frac {1\!+\!v(\eta)}{2\left[\hsU(\eta,\!z)\right]^2}-
\frac{\xi}2, 
\ea          \label{sDelta1Def}    
\eea
$
\tilde\xi_2^{{\scriptscriptstyle (1)}}(Z)$ is well-defined because $
f(\xi,\!Z)$ has a unique maximum. In \cite{FioDeAFedGueJov22} we have also determined upper, lower bounds for
$\hH(\xi,\!Z)$.
 We stress that $\check n(\xi,Z)$ is  the {\it initial} (not the present) density $\widetilde{n_0}(z)$ at $z=\hze(\xi,\!Z)$. The meaning of $n_u,n_d$ is illustrated in fig. \ref{fig1}.c.
If  the pulse is SS then: in (\ref{Deltabound1}) $\hDU(\xi,Z)$ can be replaced by $0$;
in (\ref{nbound1}) $n_u,n_d$ can be replaced by $n_u'',n_d''$, where $n_d''(z)\equiv\min\limits_{z\le\zeta\le z\!+\!\hDOu}\big\{\widetilde{n_0}(\zeta)\big\}$.

\medskip
From (\ref{Deltabound1}),  (\ref{sbound1}) we obtain also some apriori sufficient conditions for the pulse to be
SS, ES. \ $\hDU(\xi,Z)=d(\xi,Z)=f'(\xi,Z)$ vanishes at $\xi=0$, 
grows up to its unique positive  maximum at $\tilde\xi_1^{{\scriptscriptstyle (1)}}$,
then decreases to negative values; $\tilde\xi_2^{{\scriptscriptstyle (1)}}$ is the unique 
$\xi>\tilde\xi_1^{{\scriptscriptstyle (1)}}$ such that $\hDU(\xi,Z)=0$. Hence, 
$\tilde\xi_2^{{\scriptscriptstyle (1)}}$ is a lower bound for 
 $\tilde\xi_2$. Therefore the condition \ $\hDU(l,Z)\ge 0$ \
ensures that $\tilde\xi_2(Z)\ge\tilde\xi_2^{{\scriptscriptstyle (1)}}(Z)\ge l$, i.e.
the pulse is SS. Similarly,  $\hsD\!-\!1$ vanishes at $\xi=0$, 
grows up to its unique positive positive maximum at $\tilde\xi_2^{{\scriptscriptstyle (1)}}$,
then decreases to negative values. Hence, 
a lower bound $\tilde\xi_3^{{\scriptscriptstyle (1)}}$ for  $\tilde\xi_3$   is the unique 
$\xi>\tilde\xi_2^{{\scriptscriptstyle (1)}}$ such that $\hsD(\xi,Z)=1$, and 
$\tilde\xi_3(Z)\ge\tilde\xi_3^{{\scriptscriptstyle (1)}}(Z)\ge l\equiv \tilde\xi_3'$, namely that 
the pulse is ES, if 
\be
\hsD(l,Z)\ge 1.                                   \label{ShortPulse1'}
\ee 


\bigskip
{\bf Constant  initial density}. \
If 	$\widetilde{n_0}(Z)\!\equiv\!n_0$, 
then \ $\U(\Delta)\!=\!M\Delta^2/2$,  $s'=M\Delta$,
where $M\!
\equiv \!Kn_0$.
In fig. \ref{graphs} 
we plot  a monochromatic laser pulse  slowly modulated by a Gaussian and 
 the corresponding solution $(s,\Delta)$. 
The qualitative behaviour of the solution remains the same also
if $\widetilde{n_0}(z)\!\neq$ const. 
The above functions simplify:
\bea
&& \sU(\xi;\!M)= 1 +\frac {M}{2}\!\int_0^\xi\!\!\!\!d\eta\,(\xi-\eta)\,v(\eta),\\[-2pt]
&& \hDU(\xi;\!M)=d(\xi;\!M)=\int_0^\xi\!\!\!\!d\eta\,\frac {1\!+\!v(\eta)}{2\left[\sU(\eta)\right]^2}-
\frac{\xi}2,\\[-2pt]
&& \sD(\xi;\!M)-1=\frac{M}2\, f(\xi;\!M)=\frac{M}2\!\int_0^\xi\!\!\!\!d\eta\,\frac {(\xi\!-\!\eta)[1\!+\!v(\eta)]}{\left[\sU(\eta;\!M)\right]^2}- \frac{\xi^2}2. \label{sD-n_0}
\eea
Hence, a lower bound  $\tilde\xi_3^{{\scriptscriptstyle (1)}}$
for $\tilde\xi_3$ is the smallest $\xi>0$  such that \ $f(\xi)=0$, \ and the sufficient condition (\ref{ShortPulse1'}) ensuring that the pulse is ES boils down to \ $f(l)\ge 0$.

\section{Hydrodynamic regime up to wave-breaking}
\label{Jbound}

As said, the map  \ $\hat \bx_e(\xi,\cdot)\!:\!\bX\!\mapsto\!  \bx$ \   
is invertible,  and the HR 
is justified,  as long as 
\bea
\hat J \equiv \left|\frac{\partial\hat \bx_e}{\partial \bX}\right| \!=\!\frac{\partial\hat z_e}{\partial Z} \!=\! 1+\varepsilon> 0,\qquad \qquad 
\varepsilon \equiv \frac{\partial\hat \Delta}{\partial Z}. 
\eea
If $\hat J(\xi,\!Z)\!\le\!0$ then  $\hat z_e(\xi,\!Z')=\hat z_e(\xi,\!Z)$ for some $Z'\!\neq\! Z$, i.e. the layer of $Z'$ electrons crosses the layer of $Z$ electrons, and WB takes place.
Let $\kappa \equiv (1\!+\!v)/\hat s^3$.
Differentiating (\ref{heq1}) with respect to (w.r.t.) $Z$ we find that $\varepsilon$,
$\sigma\equiv \partial\hat s/\partial Z$
fulfill the Cauchy problem
\bea
\ba{ll}
\varepsilon'=-\kappa\sigma, \qquad 
&\sigma'=K\left(
\check n -\! \widetilde{n_0}+\check n\,\varepsilon\right), \\[6pt]
\varepsilon(0,Z)=0,\qquad &\sigma(0,Z)=0,  
\ea          \label{basic''} 
\eea
Differentiating the periodicity identity  $\hat z_e\!\left[\xi\!+\!n\xiH(\!Z\!),\!Z\right]\!=\!\hat z_e(\xi,\!Z)$  w.r.t. $Z,\xi$ yields  \cite{FioDeNAkhFedJov23}
\be
\hat J\!\left(\xi\!+\!n\xiH,Z\right)=\hat J(\xi,Z)\!-\!n\frac{\partial\xiH}{\partial Z}\Delta'\!\left(\xi,Z\right)\!,
 \qquad     \forall \:\: \xi\!\ge\! l,\:\:  n\in\NN,\:\:  Z\ge 0, 
 \label{pseudoper}
\ee
so that (\ref{lin-pseudoper}) holds with   $b\equiv -\hat\Delta' \frac{\partial \log\xiH}{\partial Z}$, 
$a\equiv \hat J\!-\!\xi b$. This is consistent \cite{FioDeNAkhFedJov23}
 with Floquet theorem applied to (\ref{basic''}).
Known $ \hat J,\sigma$ in $[l, l\!+\!\xiH[$ we extend them to all $\xi\ge l$ via (\ref{pseudoper}).

\subsubsection*{Bounds on $\hat J$ for small $\xi>0$, and no-WBDLPI conditions} 

 To bound $\varepsilon,\sigma$  for small $\xi$ we introduce the Liapunov function  
\be
V\equiv  \varepsilon^2\!+\!b\,\sigma^2,\qquad b\equiv 1/M_ul^2.
\ee
Using $|\varepsilon|\le\sqrt{V}$, $V(0,\!Z)=0$,
(\ref{basic''}), the Comparison Principle \cite{Yos66} one shows 
 \cite{FioDeAFedGueJov22}  that 
\bea
|\varepsilon(\xi,\!Z)| &   \!\!\le\!\! & \delta(Z)  \sqrt{\!M_u(Z)}\!\!
\int^l_0\!\! d\eta\, \exp\!\left\{\!\frac{\sqrt{\!M_u(Z)}}2\left[ (l\!-\!\eta)\, \delta(Z)
\!+\!\! \int^l_\eta\!\!\!\!d\zeta\, \D(\zeta,\!Z) \right]\!\right\} =:Q_2(Z)
 \label{NoWBcond2}\nn
&   \!\!\le\!\! & \delta(Z)\sqrt{\!M_u(Z)}  \!\!\int^l_0\!\!\!d\eta\, \exp\!\left\{\!\frac{\sqrt{\!M_u(Z)}}2\left[\!(l\!-\!\eta)\, \delta(Z) \!+\!\! 
\int^l_\eta\!\!\!d\zeta\, \tilde v(\zeta) \!\right]\!\!\right\} \!=:\! Q_1(Z)\quad \label{NoWBcond1} \nn
&  \!\!\le\!\! &\frac{2\delta(Z)}{\tvM\!+\!\delta(Z)} \left\{\exp\left[\frac{\tvM\!+\!\delta(Z)}2\sqrt{\!M_u(Z)}\,l\right]-1\right\}=:Q_0(Z) \qquad\forall\xi\in[0,\tilde\xi_3'],   \label{NoWBcond0} \nonumber
\eea
\vskip-.5cm
\bea
\mbox{where}\quad &\displaystyle \delta(Z)\equiv  1-\frac{n_d(Z)}{n_u(Z)},\quad
&\D(\xi,\!Z)  \!\equiv \! \max\left\{ \frac{1\!+\!  v(\xi)}{\left[\hsD(\xi,\!Z)\right]^3} - 1 \:, \:
1 - \frac{1\!+\!  v(\xi)}{\left[\hsU(\xi,\!Z)\right]^3}\right\}\nn
&&  \:\:\le\!\!  \tilde v(\xi) \, 
\equiv  \, \max\{v(\xi),1\} \le \max\{\vM,1\}\, =: \,\tvM; \nonumber
\eea
$\delta$ is the maximal relative variation of  $\widetilde{n_0}(Z)$ across the interval
 $[Z\!+\!\hDd,Z\!+\!\hDOu]$ swept by $\hze(\xi,Z)$ for $0\le\xi\le\tilde\xi_3'$. \ \
Consequently, \ if (\ref{ShortPulse1'}) and either $Q_0(Z)<1$, or $Q_1(Z)<1$,  or  
$Q_2(Z)<1$ are satisfied  for all $Z$, then there is no  WBDLPI
(Theorem 1  in \cite{FioDeAFedGueJov22}). Since $Q_0(Z)<1$ 
if \ $K n_b l^2<4\big[\log 2/(1\!+\!\tvM)]^2$, \  to exclude WBDLPI it suffices that 
 \  $K n_b l^2<\min\left\{4\big[\log 2/(1\!+\!\tvM)]^2,2/(1\!+\!2\Delta_u/l)\right\}$ \
(Corollary 1  in \cite{FioDeAFedGueJov22}).

\medskip
In the NR regime $v\ll 1$,  whence $\hs\simeq1$,
  $\kappa\simeq1$, $\hDOu\ll l$,  $\D\simeq 0$. If $\D$ can be neglected w.r.t. $\delta$, then $Q_2(Z)<1$ reduces to
\be
r(Z)
\equiv \delta(Z)\,\sqrt{\!Kn_u(Z)}\,l\: < \: 0.81
. \label{NR-NoWBcond}
\ee
[because $r\simeq 0.81$ makes $2\big(e^{r/2}-1\big)$ equal 1]. \
To exclude WBDLPI (\ref{NR-NoWBcond}) must be satisifed for all $Z$. This is automatically the case if $G_b<0.81/2\pi$, because $\delta\le 1$  by definition and the pulse is ES [cf. (\ref{Lncond})]. Otherwise  it is a rather mild  condition on $\delta$. \
It is sufficient to check  (\ref{NR-NoWBcond})  at maximum points of $r(Z)$.
We assume that $\widetilde{n_0}(Z)$ is piecewise continuous,
and a $Z$-independent $\hDd$, e.g. $\hDd=-\hDOu$. We now prove

\medskip
\begin{proposition}
$r(z)$ decreases (resp. grows) with $z$ in $\I=[z_1, z_2]$  ($z_2>z_1$) if in $\I'\equiv [z_1\!+\!\hDd, z_2\!+\!\hDOu]$ $\widetilde{n_0}(z)$  is concave and grows (resp. decreases)  with $z$.
\label{prop1}
\end{proposition}

\smallskip
{\bf Proof}. \ \  We recall that $f$ is concave iff  \ 
$f\big[(1\!-\!t)x\!+\! ty\big]\ge (1\!-\!t)f(x)\!+\! tf(y)$ \ for all 
$x,y$ and $t\in[0,1]$.
The claim holds iff it does for $w(z)\equiv r(z)/l\sqrt{K}$. Clearly,
\bea
w(z)-w(z') = \frac{n_u(z)\!-\!n_d(z)}{\sqrt{n_u(z)}}-\frac{n_u(z')\!-\!n_d(z')}{\sqrt{n_u(z')}}
\eea
$z\in\I$ implies $z\!+\!\hDd,z\!+\!\hDOu\in\I'$; assume $z,z'\in\I$ with $z<z'$. \
If in $\I'$ $\widetilde{n_0}(z)$ grows, then:
$n_u(z)\!=\!\widetilde{n_0}(z\!+\!\hDOu)$,  $n_d(z)\!=\!\widetilde{n_0}(z\!+\!\hDd)$; similarly for $z'$; $n_u(z)\!\le\! n_u(z')$, and
\bea
w(z)-w(z') & \ge & 
\left[
n_u(z)-n_d(z) -n_u(z')+n_d(z')\right]/ \sqrt{n_u(z)}    \nn[6pt]
& = & \frac{\widetilde{n_0}(z\!+\!\hDOu)+\widetilde{n_0}(z'\!+\!\hDd)-\widetilde{n_0}(z\!+\!\hDd)-\widetilde{n_0}(z'\!+\!\hDOu)}{\sqrt{n_u(z)}}\label{inter1}
\eea
We set $x=z\!+\!\hDd$, $y=z'\!+\!\hDOu>x$, and look for $t,u$ such that 
\be
\ba{l}
z\!+\!\hDOu=(1\!-\!t)x\!+\! ty=(1\!-\!t)(z\!+\!\hDd)\!+\! t(z'\!+\!\hDOu), \\
 z'\!+\!\hDd=(1\!-\!u)x\!+\! uy=(1\!-\!u)(z\!+\!\hDd)\!+\! u(z'\!+\!\hDOu).
\ea    \label{combi}
\ee
The solutions \
$t=\frac{\hDOu\!-\!\hDd}{z\!-\!z'\!+\!\hDOu\!-\!\hDd}$, \
$u=\frac{z\!-\!z'}{z\!-\!z'\!+\!\hDOu\!-\!\hDd}$ \
belong to $[0,1]$ and fulfill $ t\!+\!u=1$.
Replacing in (\ref{inter1}) and using the concavity of $\widetilde{n_0}$ we find 
\bea
w(z)-w(z') & \ge &  \frac{t\!+\!u\!-\!1}{\sqrt{n_u(z)}}\big[\widetilde{n_0}(y)-\widetilde{n_0}(x)\big]=0,
\label{inter2}
\eea
which shows that $w(z)$ decreases with $z$, as claimed.
\ Otherwise, if in $\I'$ $\widetilde{n_0}(z)$ decreases, then: $n_d(z)\!=\!\widetilde{n_0}(z\!+\!\hDOu)$,  $n_u(z)\!=\!\widetilde{n_0}(z\!+\!\hDd)$; similarly for $z'$; $n_u(z)\!\ge\! n_u(z')$, and
\bea
w(z)-w(z') & \le & \frac{1}{\sqrt{n_u(z')}}\left[
n_u(z)-n_d(z) -n_u(z')+n_d(z')\right]\nn[6pt]
& = & \frac{\widetilde{n_0}(z\!+\!\hDd)+\widetilde{n_0}(z'\!+\!\hDOu)-\widetilde{n_0}(z\!+\!\hDOu)-\widetilde{n_0}(z'\!+\!\hDd)}{\sqrt{n_u(z')}}\label{inter1'}
\eea
Imposing again (\ref{combi}), we find the same solutions $t,u$.
Replacing in (\ref{inter1'}) and using the concavity of $\widetilde{n_0}$ we find 
that $w(z)$ grows with $z$, as claimed, because
\bea
w(z)-w(z') & \le &  \frac{t\!+\!u\!-\!1}{\sqrt{n_u(z)}}\big[\widetilde{n_0}(x)-\widetilde{n_0}(y)\big]=0.\qquad\qquad\mbox{QED}
\label{inter2'}
\eea

\begin{corollary}
In a concavity interval $\I'\equiv [z_1\!+\!\hDd, z_2\!+\!\hDOu]$ of $\widetilde{n_0}(z)$, 
(\ref{NR-NoWBcond}) is satisfied at all $z\in\I=[z_1, z_2]$ if: i) it is at $z=z_1$, in the case
$\widetilde{n_0}(z)$ grows in all of $\I$;  ii) it is at $z=z_2$, in the case
$\widetilde{n_0}(z)$ decreases in all of $\I$;  iii) it is at $z=z_1,z_2$, otherwise.
\label{coroll}
\end{corollary}

\medskip
\noindent
In other words, under the above assumptions the maximum points(s) of $r(z)$ in $\I$ are one or both extremes, while they can be  also inside $\I$  if $\widetilde{n_0}(z)$ is convex in $\I'$.  If $\widetilde{n_0}\in C^1(\I')$ they satisfy the equation (which has no solution if $\widetilde{n_0}$ is concave in $\I'$)
\be
w'(z)=0 \quad\Leftrightarrow\quad n_d'(z)=n_u'(z)\frac{1+n_d(z)/n_u(z)}2.
\label{interNR}
\ee

\section{Discussion and conclusions}
\label{discuss}

We have formulated the conditions for ES, SS pulses and no WBDLPI in terms of
dimensionless functions ($v,g,\sU,\sD,\, M_ul^2,\, M_dl^2,\, \delta,$...) and  numbers ($\tvM,\, \Delta_u/l,\, 
G_b^2,$...) characterizing the input data. One can compute these quantities and  check
the conditions in few seconds running a specifically 
designed program  that uses some general-purpose numerical package (like {\it Mathematica}) on a common notebook. Often one can check the conditions just by a 
back-of-the-envelope estimate of these quantities. A rescaling of the input data
that leaves these dimensionless quantities invariant does not affect the fulfillment
of the conditions. 

\begin{figure}[t]
\includegraphics[height=12.5cm]{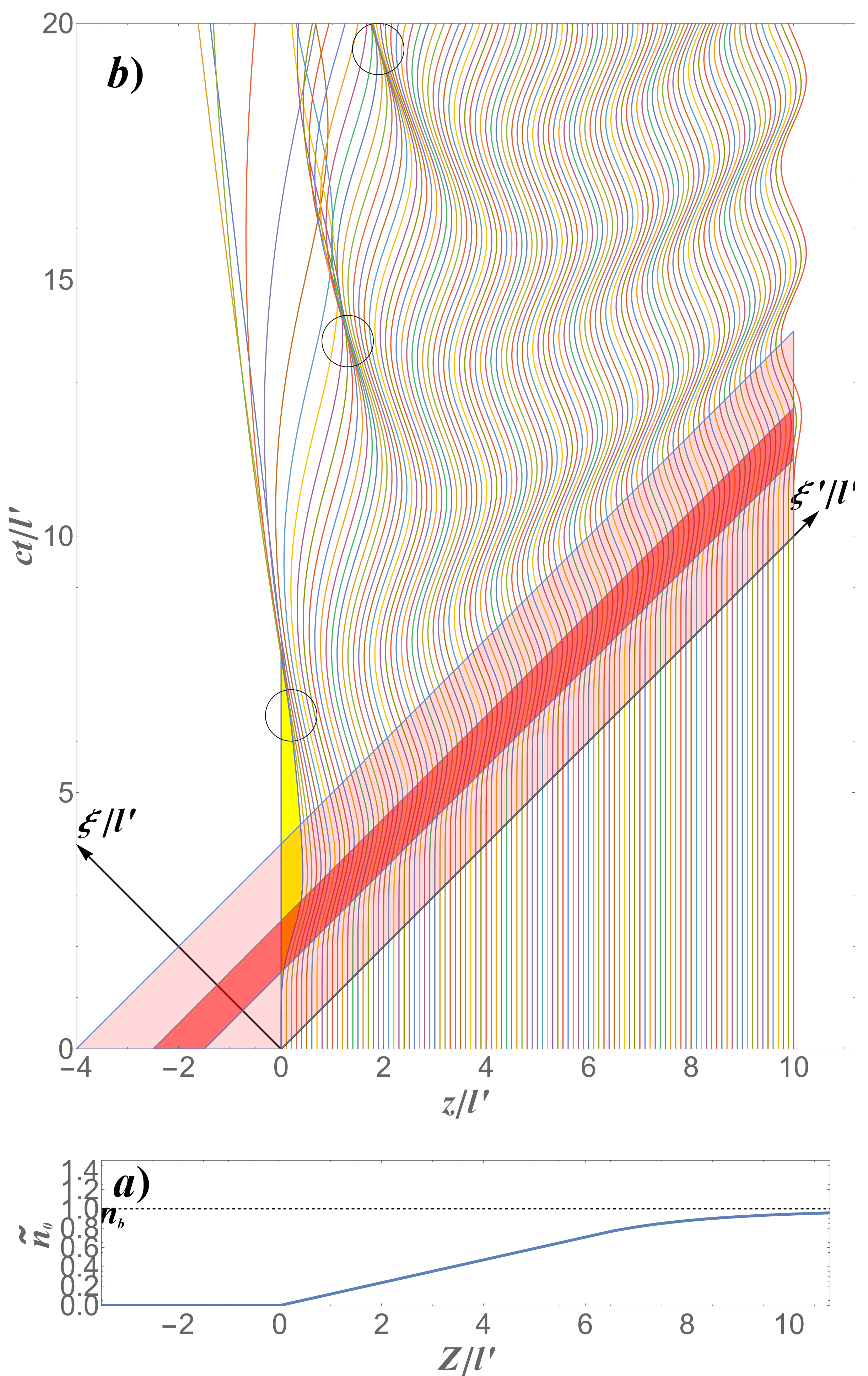}\hfill
\includegraphics[height=12.5cm]{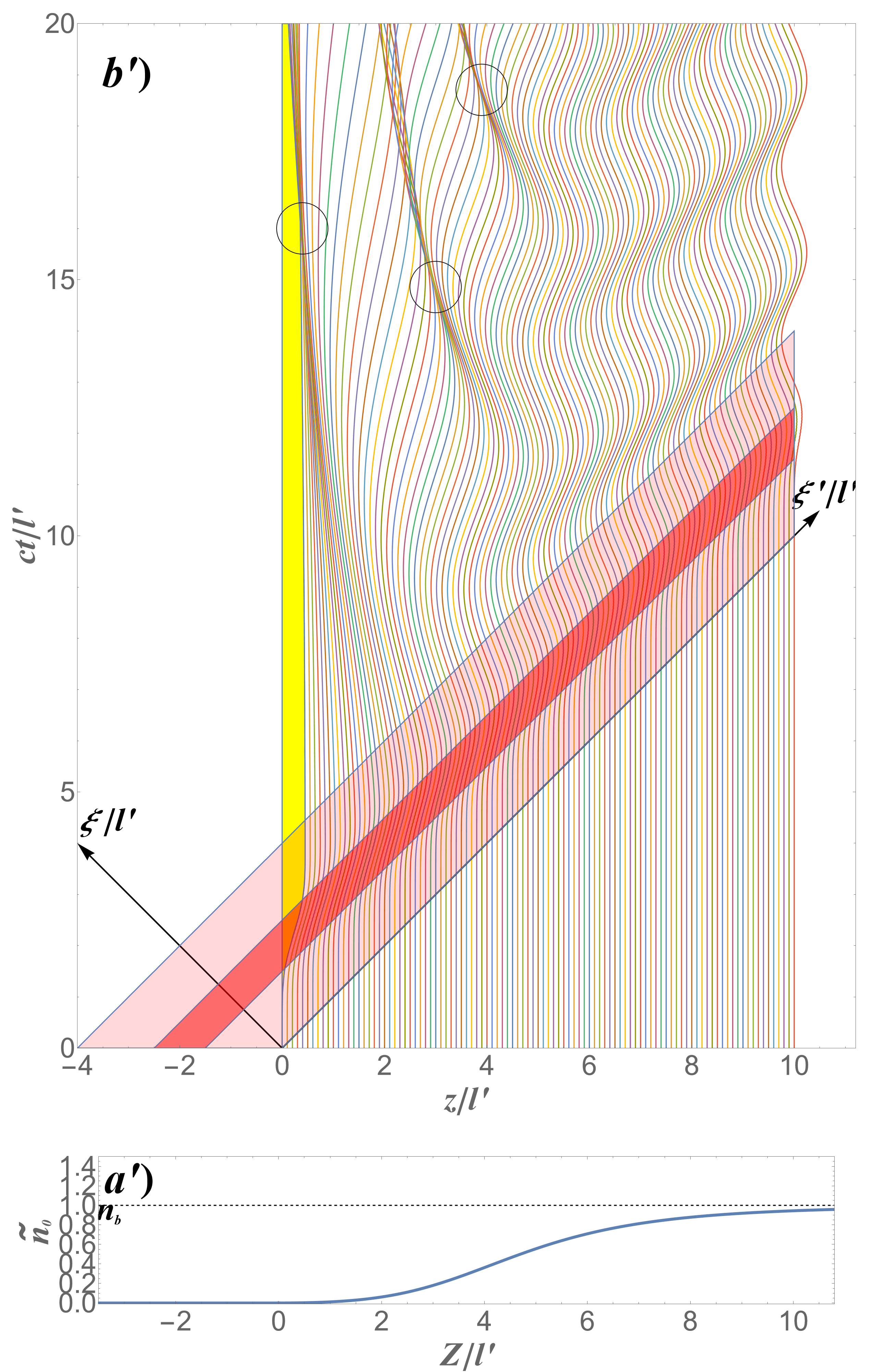}
\caption{Monotonic $\widetilde{n_0}$s 
sharing the 
asymptotic value $n_b\!=\!4/Kl'{}^2$ and: a)  $\widetilde{n_0}(z)\!=\! O(z)$, a') $\widetilde{n_0}(z)\!=\! O(z^2)$ [cf. (\ref{linear}),  (\ref{quadratic})] hit by the ES pulse of fig. \ref{graphs}.  \ b), b'): Corresponding projections onto the $z,ct$ plane of the  WLs of the $Z\!=\!0,\!2\lambda,\!...,\!200\lambda$ electrons; encircled are the earliest WBs (crossing WLs); $\xi'\!\equiv\! ct\!+\!z$. The support $0\!\le\! ct\!-\!z\le l$  of $\Bep(ct\!-\!z)$ is pink; the region $(l\!-\!l')/2\!\le\!  ct\!-\!z\!\le\!  (l\!+\!l')/2$ where the modulating intensity 
exceeds half maximum is red; the pure-ion layer spacetime region is  yellow.}
\label{Worldlines_HomLin-Nonlin}
\end{figure}

To shed some light on these conditions we assume for simplicity a continuous $\widetilde{n_0}$ and consider first  the NR regime. 
Given $n_b,\bar Z>0$, which $\widetilde{n_0}$ grow up to  $n_b$ in $[0,\bar Z]$, 
but do not cause WBDLPI? In particular, which one(s) do with the least $\bar Z$?

If the growing $\widetilde{n_0}(z)$ is concave for $z>0$, then to avoid WBLDPI it suffices that 
$r(0)=l\sqrt{K\widetilde{n_0}(\Delta_u)}<0.81$, say $r(0)=0.8$, by Corollary \ref{coroll}. We minimize $\bar Z$ by the steepest concave $\widetilde{n_0}(z)$ yielding $r(0)=0.8$, i.e. the straight half-line 
\be
\widetilde{n_0}(z)=n_1(z)\equiv z\,\theta(z) \: 0.64/Kl^2\Delta_u;
\label{linear}
\ee
 then
$\widetilde{n_0}\big(\bar Z\big)=n_b$ with $\bar Z=\bar Z_1\equiv Kn_bl^2\Delta_u/0.64$.
If the growing $\widetilde{n_0}(z)$ is convex for $z>0$ we minimize $\bar Z$ by imposing $r(z)=0.8$ for all $z\in[0,\bar Z]$, whence $w'(z)=0$ identically; the  solution of this equation depends  on the specific values $\Delta_u,\Delta_d$. In the limit $\Delta_u,\Delta_d\to 0$ (very NR limit), if $\widetilde{n_0}\in C^2(]0,\bar Z[)$,   by (\ref{interNR}) this becomes
$$
\frac {\widetilde{n_0}''}{\widetilde{n_0}'}=\frac {\widetilde{n_0}'}{2\widetilde{n_0}}
\quad\Leftrightarrow\quad \frac d{dz} \log \left[\frac {\sqrt{\widetilde{n_0}}}{dz}\right] (z)=0;
\label{interNR'}
$$
its solutions have the form $\widetilde{n_0}(z)=\theta(z) (cz\!+\!d)^2$; imposing
$\widetilde{n_0}(0)\!=\!0$, $\widetilde{n_0}\big(\bar Z\big)\!=\!n_b$ yields 
\be
\widetilde{n_0}(z)=n_2(z)\equiv \theta(z) n_b\big(z/\bar Z\big)^2.
\label{quadratic}
\ee 
It is easy to check that $r'(z)>0$, so that
$r(z)$ grows for all $z>0$. To avoid WBLDPI  it suffices that $r\big(\bar Z\big)=0.8$; assuming $\Delta_d=-\Delta_u$, by a little algebra this leads to
\be
0.8=r\big(\bar Z\big)\le \sqrt{Kn_b}l \: 4\Delta_u/\bar Z \quad\Rightarrow\quad 
\bar Z\le\bar Z_2\equiv 5\sqrt{Kn_b}l \,\Delta_u.
\ee
Hence, \ $\bar Z_1/\bar Z_2=0.128\sqrt{Kn_b}l\simeq 0.8 G_b$, \
and  the linearly (resp. quadratically) growing density (\ref{linear}) [resp. (\ref{quadratic})] is preferable if $G_b\le 1.125$ [cf. (\ref{Lncond})] (resp. if $G_b> 1.125$). This is confirmed e.g. by fig. \ref{Worldlines_HomLin-Nonlin}: the earliest WB 
occurs at a much smaller $\xi$ with the density of type a)  than 
with that of type a'), and $Q_2>Q_2'$, e.g. $Q_2(l'/2)\simeq 2$,
$Q_2'(l'/2)\simeq 0.7$.

\bigskip
 In LWFA experiments $G_b$ may considerably exceed 1 even
with ES pulses leading to   moderately relativistic regimes.
Again, growing quadratic densities (\ref{quadratic}) prevent WBDLPI by a smaller $\bar Z$ than linear ones (\ref{linear}).  
In the most typical LWFA experiments  one shoots a laser pulse orthogonally to a supersonic diluted gas (e.g. hydrogen or helium) jet coming out of a nozzle in a vacuum chamber; the jet is ionized into a plasma by the front of the pulse. Correspondingly,
$\widetilde{n_0}(0)=0=\frac{d\widetilde{n_0}}{dZ}(0)$ (see e.g. fig. 2 in \cite{HosEtAl02}), 
and $\widetilde{n_0}(z)= O(z^2)$, i.e.  for small $z>0$ the density  is  typically convex and closer to  type  (\ref{quadratic})  than to type  (\ref{linear}), and thus more easily prevents WBDLPI. For larger $z$  the density becomes concave and tends to an asymptotic value $n_b$; assuming Proposition  \ref{prop1} and Corollary \ref{coroll} keep valid
in these regimes, the fulfillment of either $Q_0(Z)<1$, or $Q_1(Z)<1$,  or  
$Q_2(Z)<1$ at the inflection point (which is the right extreme of the convexity interval and the left extreme of the concavity one) is a strong indication that no WBDLPI takes place (what can be checked by solving (\ref{basic''}) numerically).

If more realistically the pulse is not a plane wave, but cylindrically symmetric around  $\vec{z}$  with a {\it finite} spot radius $R$ (which we assume to stay constant in the plasma, by self-focusing), then - by causality - our results hold strictly inside the causal cone (of axis $\vec{z}$, radius $R$) trailing the pulse, and approximately in a neighbourhood thereof, as far as the pulse is not significantly affected by its interaction with the plasma;  for typical LWFA experiments this means travelling  many $l$ in the $z$-direction \cite{FioDeNAkhFedJov23,FioAAC22}.

\medskip
{\bf Acknowledgments.} \ 
Work done also within the activities of GNFM. 
These results have been partially presented in the International Conference WASCOM21.

\medskip

\bibliographystyle{plain}
\bibliography{RicercheMat-HydroImpact2-ArXive}

\end{document}